\documentclass[twocolumn,prb,showpacs]{revtex4}
\usepackage{graphicx}
\usepackage{dcolumn}
\usepackage{bm}
\usepackage{color}
\usepackage{mathrsfs}

\begin{document}
\title{Gapped state in three-leg $S=\frac{1}{2}$ Heisenberg tube
}
\author{S.~Nishimoto}
\affiliation{IFW Dresden, Leibniz-Institut f\"ur Festk\"orper- und Werkstoffforschung 
Dresden, D-01171 Dresden, Germany}
\author{M.~Arikawa}
\affiliation{Institute of Physics, University of Tsukuba 1-1-1 Tennodai, Tsukuba Ibaraki 305-8571, Japan}
\date{\today}
\begin{abstract}
We study the ground state of three-leg $S=\frac{1}{2}$ Heisenberg tube using 
the density-matrix renormalization group method. The dimerization order-parameter 
and spin-excitation gap are calculated in a wide range of leg exchange 
interactions. We confirm that a gapped state with the dimerization order 
is realized even when the leg exchange interactions are ferromagnetic. 
Furthermore, the topological configuration of spin-singlet pairs in the ordered 
state is determined from the results of the Berry phase of each spin coupling 
as well as the structure factor of singlet-singlet correlation functions. 
We find that there exist three kinds of configurations of the valence-bond states 
depending on the ratio of leg and rung exchange interactions.
\end{abstract}
\maketitle 

Odd-leg spin ladder belongs to the same universality class as 
single chain; thus, the ground state is comprehended as a gapless 
spin-liquid (or a Tomonaga-Luttinger liquid).~\cite{Dagotto96} 
However, if the periodic boundary conditions are applied in the rung 
direction, i.e., a tube is shaped, the spin states are dramatically 
changed. It is due to the occurrence of a geometric property called 
{\it frustration}, which is today a hot topic in condensed matter 
physics.~\cite{Moessner06} In general, it would appear that the fundamental 
low-energy physics of any odd-leg spin tube is essentially epitomized by 
that of three-leg one. So far, it has been recognized that, as long as 
all the exchange interactions are antiferromagnetic, the three-leg spin 
tube can be spontaneously dimerized to avoid (or to reduce) the frustration 
and the spin excitations are gapped.~\cite{Schulz96,Kawano97}

An ideal nanotubular material with odd number of legs is vanadium oxide 
Na$_2$V$_3$O$_7$, which may be regarded as a $S=\frac{1}{2}$ nine-leg 
Heisenberg spin tube system.~\cite{Millet99} In experiments (Ref.~\onlinecite{Gavilano05}), 
the $^{23}$Na NMR response, the dc- and ac-magnetic susceptibilities, 
and the specific heat reveal that above 100~K the system is considered 
as paramagnetic; whereas, below 100~K most of the localized V magnetic 
moments ($S=\frac{1}{2}$) form a collection of spin-singlet dimers with 
gaps $\Delta \sim 0-350$~K and the remaining small fraction of them 
forms spin-triplet bound states with gaps $\Delta \sim 0-15$~K; and, 
the degeneracy of the triplet ground states is lifted by a phase transition 
at 0.086 K. The mechanism of the gap opening is still open issue.
Besides, this material has attracted considerable attention 
from a standpoint of entanglement in spin systems.~\cite{Toth09,Souza09} 
Moreover, it is of great interest to seek a relevance to a spin-liquid 
state observed in a three-leg $S=\frac{3}{2}$ spin tube system CsCrF$_4$.~\cite{Manaka09}

The low-energy spin Hamiltonian of Na$_2$V$_3$O$_7$ has been proposed 
by some theoretical groups, nevertheless, it is still controversial. Both 
the rung ($J_\perp$) and leg ($J_\parallel$) exchange interactions 
seem to be very sensitive because the estimated values are quite different, 
even in ferro- and antiferromagnetic characteristics, depending on 
the approaches. An overview of the results are as follows. 
(i) The {\it ab initio} microscopic analysis:~\cite{Dasgupta05} 
both $J_\perp$ and $J_\parallel$ are antiferromagnetic, but $J_\parallel$ 
is frustrated and the magnitude is much smaller than $J_\perp$; 
(ii) the first-principle calculations:~\cite{Mazurenko06} $J_\perp$ is 
antiferromagnetic, while $J_\parallel$ is ferromagnetic, and they have 
the same order of magnitude; and, (iii) the first-principles density 
functional theory:~\cite{Fortea09} $J_\parallel$ is ferromagnetic and 
$J_\perp$ is ferro- or antiferromagnetic. Possibly, the point to be 
grasped next is whether the dimerization order with finite spin gap 
can occur when ferromagnetic exchange interactions are contained in 
the odd-leg spin tube.

We thus consider the ground state and low-lying excited states of three-leg 
$S=\frac{1}{2}$ Heisenberg spin tube and provide new insights especially 
for the case that the ferromagnetic exchange interactions are taken 
into account. The Hamiltonian is given by
\begin{eqnarray}
H = J_\parallel \sum_{\alpha=1}^3 \sum_{i=1}^L \vec{S}_{\alpha,i} \cdot \vec{S}_{\alpha,i+1} 
+ J_\perp \sum_{\alpha (\neq \alpha^\prime)} \sum_{i=1}^L \vec{S}_{\alpha,i} 
\cdot \vec{S}_{\alpha^\prime,i},
\label{hamiltonian}
\end{eqnarray}
where $\vec{S}_{\alpha,i}$ is a spin-$\frac{1}{2}$ operator at rung $i$ and 
leg $\alpha$. We here restrict the rung interactions $J_\perp$ to be antiferromagnetic 
because it is obvious that no dimerization order occurs with ferromagnetic 
rung interactions. The leg interactions $J_\parallel$ are varied from ferro- to 
antiferromagnetic range. Although several theoretical studies have been carried out 
on this and similar models, only antiferromagnetic interactions are considered.~\cite{Sakai05,Cabra97,Citro00,Sato07,Luscher04,Okunishi05,Fouet06,Nishimoto08}. 

In this paper, we calculate the dimerization order-parameter and the spin-excitation 
gap to clarify in which range of the exchange interactions the ordered state 
is realized. Also, the Berry phase of each coupling and the structure factor of 
singlet-singlet correlation functions are calculated to check topological configuration 
of the spin-singlet pairs in the ordered state. For those calculations, the density-matrix 
renormalization group (DMRG) technique~\cite{White92} is applied. We investigate 
tubes with several kinds of length up to $L=312$, i.e., $312 \times 3$ cluster, 
under the open boundary conditions (OBC) in the leg direction, unless otherwise stated. 
The density-matrix eigenstates up to $m=2400$ are kept in the renormalization 
procedure and all quantities are extrapolated to the limit $m \to \infty$. 
In this way, the discarded weight is less than $1 \times10^{-7}$, while 
the maximum error in the ground-state energy is less than $10^{-7}-10^{-6}$.

\begin{figure}[t]
    \includegraphics[width= 6.5cm,clip]{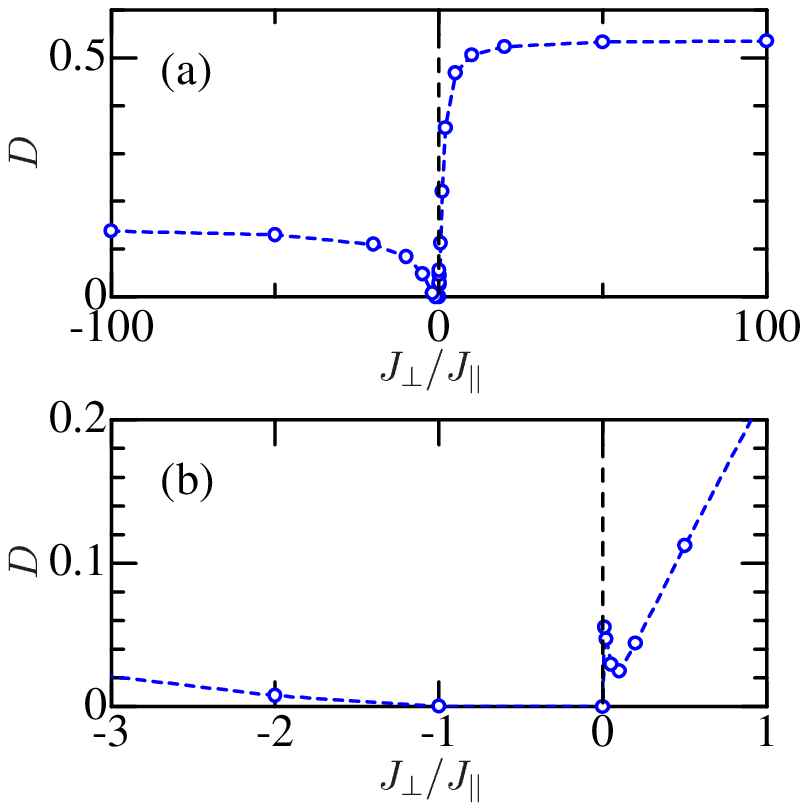}
  \caption{
(a) Dimerization order-parameter $D$ extrapolated to the thermodynamic limit 
$L \to \infty$ as a function of $J_\perp/J_\parallel$. (b) Extended figure 
of (a) for $-3 \le J_\perp/J_\parallel \le 1$.
  }
    \label{fig_OP}
\end{figure}

Let us first evaluate the dimerization order-parameter in a wide range of 
$J_\perp/J_\parallel$ for exploring the presence or absence of long-range 
dimerized state. The order parameter is featured by an alternation of 
nearest-neighbor spin-spin correlations, 
$S(i) = -\langle \vec{S}_{\alpha,i} \cdot \vec{S}_{\alpha,i+1} \rangle$,
where $\langle \cdots \rangle$ denotes the ground-state expectation value. 
As the OBC breaks translational symmetry in our calculation, 
the dimerized state is observed as a Friedel oscillation. Generally, 
the amplitude of the Friedel oscillation at system center decays as 
a function of system size. In case that it persists for arbitrarily 
large system size, we can judge that a long-range order exists. 
Thus, the dimerization order-parameter is defined as 
\begin{equation}
D = \lim_{L \to \infty} \left|S(L/2) - S(L/2+1)\right|.
\end{equation}
Nonzero value of $D$ indicates the presence of long-range dimerization 
order in our model. In Fig.~\ref{fig_OP} (a), the results of $D$ are 
shown as a function of $J_\perp/J_\parallel$. We find that 
the dimerization order appears over the entire region of 
$J_\perp/J_\parallel (\neq 0)$. For both positive and negative 
$J_\perp/J_\parallel$ values, $D$ increases rapidly at 
$J_\perp/|J_\parallel| \lesssim 10$ and keeps almost constant at 
$J_\perp \gtrsim10$: in the limit of $J_\perp/J_\parallel \to \infty$ $(-\infty)$, 
it is saturated to $D \sim 0.0673$ $(0.0183)$. However, 
quite different behaviors are seen if we take a closer look at 
small $J_\perp/|J_\parallel|$ regime [see Fig.~\ref{fig_OP} (b)]. 
At $J_\perp/J_\parallel>0$, $D$ is discontinuously enhanced when 
$J_\perp$ is switched on, then goes through a minimum around $J_\perp=0.1$, 
and increases almost linearly from $J_\perp \approx 0.2$ to $5$; 
while at $J_\perp/J_\parallel<0$, $D$ increases gradually 
with increasing $J_\perp$, like $D \sim \exp[-{\cal O}(1/J_\perp)]$. 

\begin{figure}[b]
\includegraphics[width= 8.0cm,clip]{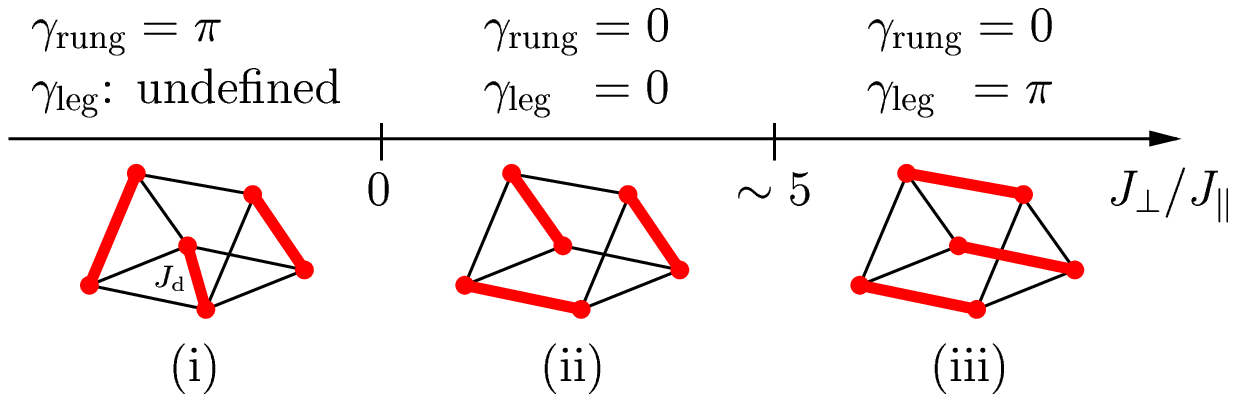}
  \caption{
Berry phases on the rung bond ($\gamma_{\rm rung}$) and leg bond ($\gamma_{\rm leg}$) 
along the parameter $J_\perp/J_\parallel$. Schematic pictures of the valence-bond 
state are also shown. Two dots linked by bold line denote a spin-singlet pair.
  }
    \label{fig_PD}
\end{figure}

The dimerization order implies a complete occupation of the system with 
spin-singlet pairs. Which poses a question on the topological configuration 
of the pairs. In order to solve it, we investigate the quantized Berry phase 
in the ordered state. The Berry phase is defined by
\begin{equation}
\gamma=-i\int^{2\pi}_{0}A(\phi)d\phi,
\end{equation}
where $A(\phi)$ is the Abelian Berry connection, 
$A(\phi)=\langle \psi_\phi | \partial_\phi \psi_\phi \rangle$ 
with the ground state $|\psi_\phi \rangle$.~\cite{Hatsugai06} The Berry phase 
is quantized as $0$ or $\pi$ (mod $2\pi$) if the system has spin gap during 
the adiabatic continuation and time reversal symmetry; and ``undefined'' 
if a gapless excitation exists. We introduce a local perturbation by a  
twist of the nearest-neighbor connection,
$
\vec{S}_{\alpha,i} \cdot \vec{S}_{\alpha^\prime,j} \to 
\frac{1}{2}(e^{-i\phi}S_{\alpha,i}^+S_{\alpha^\prime,j}^-
+e^{i\phi}S_{\alpha,i}^-S_{\alpha^\prime,j}^+)+S_{\alpha,i}^zS_{\alpha^\prime,j}^z.
$
We here pick up a dimerized pair of triangles, i.e., including six spins, 
and evaluate the Berry phases of the leg bond ($\gamma_{\rm leg}$) for 
$\alpha=\alpha^\prime, j=i+1$ and of the rung bond ($\gamma_{\rm rung}$) 
for $\alpha \neq \alpha^\prime, j=i$. Note that the dimerized pair of 
triangles must include three spin-singlet pairs. 

In Fig.~\ref{fig_PD}, the $J_\perp/J_\parallel$-dependence of $\gamma_{\rm leg}$ 
and $\gamma_{\rm rung}$ are shown and the corresponding configurations 
are also schematically described. We find that there are three kinds 
of the spin-singlet configurations along $J_\perp/J_\parallel$. 
Let us now see the case that $J_\parallel$ is antiferromagnetic. A reconstruction 
of the valence bonds is seen at $J_\perp/J_\parallel \approx 5$. 
This would correspond to the crossover between constant-$\Delta$ and 
proportional-$\Delta$ regions around $J_\perp/J_\parallel = 5$.~\cite{Nishimoto08} 
For $J_\perp/J_\parallel \gtrsim 5$, the Berry phases of leg and rung bonds 
are denoted by $\pi$ and $0$, respectively. It means that all singlet pairs 
are formed on the leg bond [Fig.~\ref{fig_PD}(iii)]. Whereas, for 
$0 < J_\perp/J_\parallel \lesssim 5$, both $\gamma_{\rm leg}$ and 
$\gamma_{\rm rung}$ are denoted by $0$. It may be interpreted if we assume 
that both the leg and rung bonds are involved to form spin-singlet pairs: 
one pair is formed in either one of three legs and the other four spins 
form a couple of pairs in the two rungs [Fig.~\ref{fig_PD}(ii)]. 
This valence-bond state is threefold degenerate, so that the spin-singlet 
pairs would be not detected as {\it local} ones. Then, we turn to the case 
that $J_\parallel$ is ferromagnetic. A valence-bond state is detected on 
the rung bond and a gapless excitation is found in the leg bond; however, 
the spin excitations are gapped in bulk limit as shown below. It will 
be settled if we suppose that two spin-singlets are formed on rungs of 
each triangle and the other one is between two sites on different legs 
and triangles [Fig.~\ref{fig_PD}(i)]. We call the last pair ``diagonal 
spin-singlet'' and the effective antiferromagnetic exchange interaction 
is denoted by $J_{\rm d}$ hereafter.

\begin{figure}[t]
    \includegraphics[width= 6.5cm,clip]{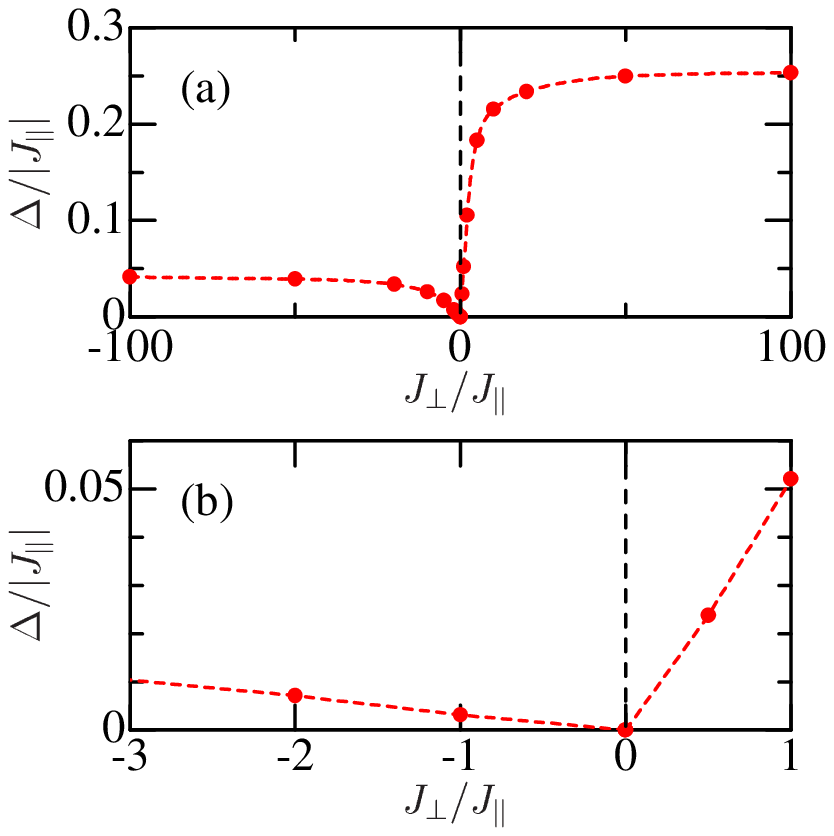}
  \caption{
(a) Spin gap $\Delta$ extrapolated to the thermodynamic limit $L \to \infty$ 
as a function of $J_\perp/J_\parallel$. (b) Extended figure of (a) for 
$-3 \le J_\perp/J_\parallel \le 1$.
  }
    \label{fig_gap}
\end{figure}

And now, of particular interest is the magnitude of spin-excitation gap 
especially in the ferromagnetic $J_\parallel$ case. It will be clarified 
by calculating an energy difference between the singlet ground state and 
the first triplet excited state,
\begin{equation}
\Delta=\lim_{L \to \infty} [E(L,1)-E(L,0)],
\end{equation}
where $E(L,S_z)$ is the ground-state energy of the system of length $L$ with
the $z$-component of total spin $S_z$. We note that the system length must 
be taken as $L=2l$, with $l (>1)$ being an integer to maintain total spin of 
the ground state as $S=0$. In Fig.~\ref{fig_gap}, we show the results of 
$\Delta$ as a function of $J_\perp/J_\parallel$. We see that the gap 
opens except at $J_\perp=0$, as anticipated from the results of dimerization 
order-parameter. For both $J_\parallel>0$ and $J_\parallel<0$, roughly 
speaking, $\Delta$ starts to increase proportionally to $J_\perp$ and 
shift into almost constant for larger $J_\perp$. It means that the lowest 
singlet-triplet excitations for small and large $J_\perp/|J_\parallel|$ 
are scaled by distinct exchange interactions. When $J_\parallel$ is 
antiferromagnetic, as suggested in our previous paper,~\cite{Nishimoto08} 
$\Delta$ is approximately scaled by $J_\perp$ ($J_\parallel$) in the small 
(large) $J_\perp/J_\parallel$ regime. On the other hand, $\Delta$ seems 
to be always scaled by $J_{\rm d}$ since the diagonal spin-singlet pair 
has the smallest binding energy when $J_\parallel<0$. It may be interpreted 
as follows: in the perturbative sense, $J_{\rm d}$ is proportional to 
$J_\perp$ ($|J_\parallel|$) for $J_\perp \ll J_\parallel$ 
($J_\perp \gg J_\parallel$); accordingly, $\Delta$ is scaled by 
$J_\perp$ ($|J_\parallel|$) in the small (large) $J_\perp/|J_\parallel|$ 
regime as in the case of antiferromagnetic $J_\parallel$. In fact, 
$\Delta$ for $J_\perp/J_\parallel<0$ is about a factor of $6$ smaller than 
that for $J_\perp/J_\parallel>0$. As a consequence, we obtain $\Delta=0.0437$ 
in the limit of $J_\perp/J_\parallel=-\infty$; $\Delta=0.254$ in the limit 
of $J_\perp/J_\parallel=\infty$. We note that the $J_\perp$-dependence of 
$\Delta$ looks similar to that of $D$ but except when $J_\perp/|J_\parallel|$ 
is very small.

\begin{figure}[t]
    \includegraphics[width= 7.0cm,clip]{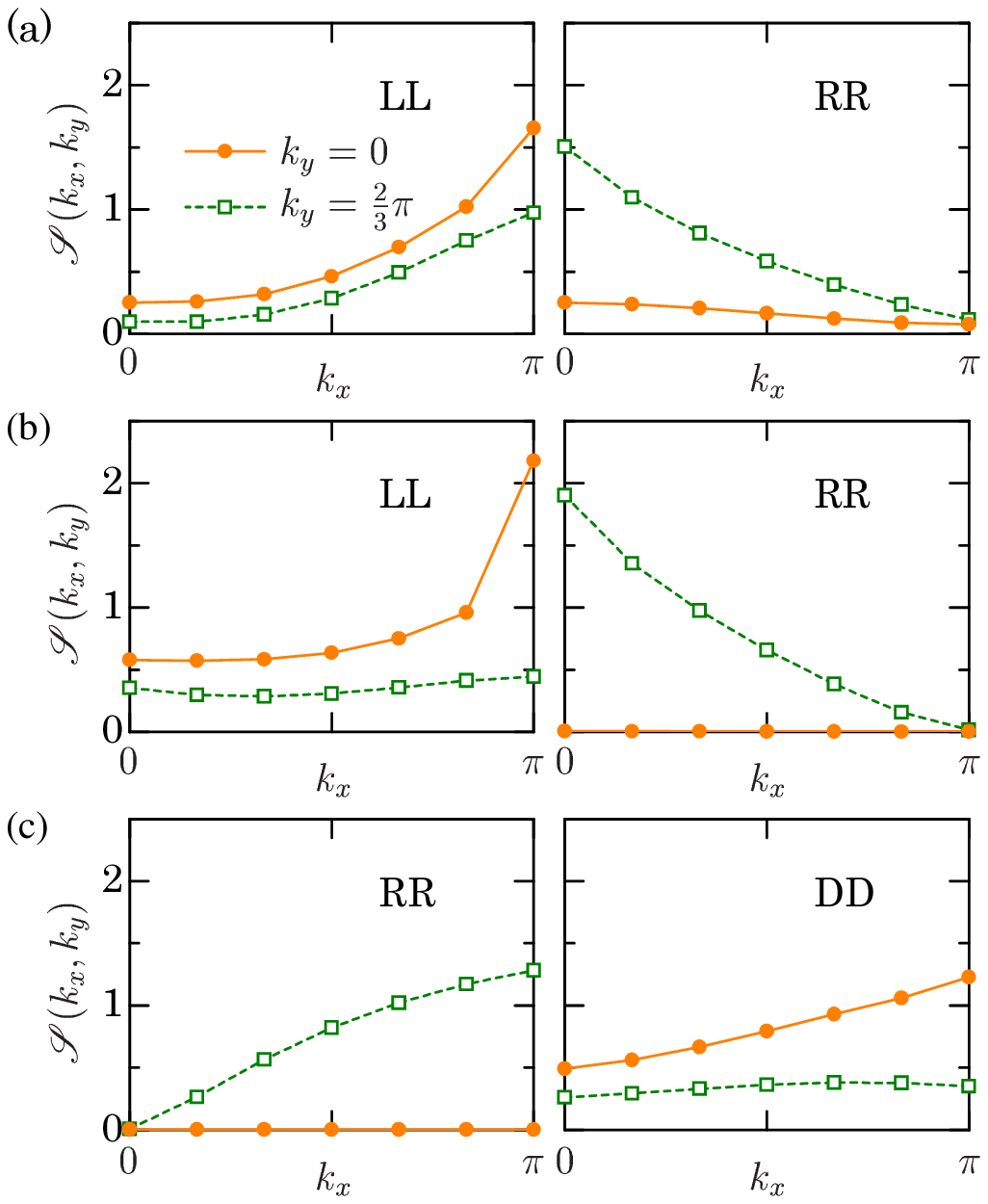}
  \caption{
Singlet-singlet correlation functions $\mathscr{S}(\vec{k})$ for 
(a) $J_\parallel=1, J_\perp=1$, (b) $J_\parallel=1, J_\perp=10$, 
and (c) $J_\parallel=-1, J_\perp=10$.
  }
    \label{fig_ss_corr}
\end{figure}

Lastly, in order to determine the periodicity of alignment of the valence-bond 
configurations, we calculate singlet-singlet correlation functions
\begin{equation}
\mathscr{S}(\vec{k})=
\frac{1}{3L} \sum_{\alpha\alpha^\prime ij} \langle \mathscr{S}_{\alpha,i} \mathscr{S}_{\alpha^\prime,j} \rangle 
\exp [i \vec{k} \cdot (\vec{r}_{\alpha,i}-\vec{r}_{\alpha^\prime,j})]
\end{equation}
with spin-singlet number operator
\begin{equation}
\mathscr{S}_{\alpha,i}=\frac{1}{4}-\vec{S}_{\alpha,i} \cdot \vec{S}_{\alpha^\prime,j},
\end{equation}
where $(\alpha^\prime,j)=(\alpha,i+1)$, $(\alpha+1,i)$, and $(\alpha+1,i+1)$ are chosen 
for the leg-leg (LL), rung-rung (RR), and diagonal-diagonal (DD) correlations, 
respectively. Here, the system size is fixed at $L=24$ using the periodic 
boundary conditions. Thus, relations $\alpha+3 \equiv \alpha$ and $i+24 \equiv i$ 
are fulfilled. 

\begin{figure}[t]
    \includegraphics[width= 6.0cm,clip]{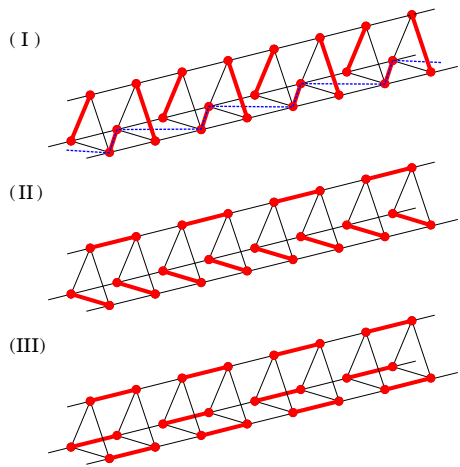}
  \caption{
Possible configurations of spin-singlet pairs for regions (I) $J_\perp/J_\parallel<0$, 
(II) $0<J_\perp/J_\parallel\lesssim5$, and (III) $5\lesssim J_\perp/J_\parallel$. 
Two dots linked by bold line denote a spin-singlet pair.
  }
    \label{fig_struc}
\end{figure}

In Fig.~\ref{fig_ss_corr}, the correlation functions $\mathscr{S}(\vec{k})$ at 
(a) $J_\parallel=1, J_\perp=1$, (b) $J_\parallel=1, J_\perp=10$, and 
(c) $J_\parallel=-1, J_\perp=10$ for $L=24$ are shown [the deviation from those 
for $L=12$ has been confirmed to be negligibly-small]. We initially point out that 
large values of $\mathscr{S}(\pi,\frac{2}{3}\pi)$ in the LL and of $\mathscr{S}(0,\frac{2}{3}\pi)$ 
in the RR correlations jointly indicate a straight link of the state (ii) in 
Fig.~\ref{fig_PD}, which is described as alignment (II) in Fig.~\ref{fig_struc}; 
whereas, large value of $\mathscr{S}(\pi,0)$ in the LL correlation suggests a link 
of state (iii), which corresponds to alignment (III). For $J_\parallel=1, J_\perp=1$, 
therefore, it would appear that the alignments (II) and (III) coexist 
[see Fig.\ref{fig_ss_corr}(a)]; however, the alignment (III) must be less dominant 
considering the results of the Berry phase. As $J_\perp$ increases, in the LL correlation 
$\mathscr{S}(\pi,\frac{2}{3}\pi)$ is rapidly diminished and $\mathscr{S}(\pi,0)$ 
is rather enhanced [see Fig.\ref{fig_ss_corr}(b)]; as a result, the alignment (III) 
becomes dominant for large $J_\perp/J_\parallel$ region. Then, we turn to the case of 
$J_\parallel<0$  [see Fig.\ref{fig_ss_corr}(c)]. As expected, $\mathscr{S}(\vec{k})$ 
in the LL correlation is always zero indicating no spin-singlet pair in the leg bond; 
instead, the DD correlations are significant. The enhancements of 
$\mathscr{S}(\pi,\frac{2}{3}\pi)$ in the RR and $\mathscr{S}(\pi,\frac{2}{3}\pi)$ 
in the DD correlations are seen. It indicates a straight alignment of the state (i), 
as shown in Fig.~\ref{fig_struc} (I). We argue that a zigzag chain, which is denoted 
as a dotted line in Fig.~\ref{fig_struc}, is dimerized to remove the frustration 
instead of being dimerized in the leg direction for antiferromagnetic $J_\parallel$.

In conclusion, we study the ground-state properties of three-leg $S=\frac{1}{2}$ 
Heisenberg tube with antiferromagnetic rung exchange interactions. Using the DMRG 
method, the dimerization order-parameter, spin-excitation gap, Berry phase, and 
structure factor of singlet-singlet correlation functions are calculated. 
We confirm that the dimerization order occurs for both ferro- and antiferromagnetic 
leg exchange interactions, and the spin excitations are always gapped. Also, 
we find that there are three kinds of phases according to topological configuration 
of the spin-singlet pairs: (I) diagonal-singlet regime for $J_\perp/J_\parallel<0$, 
(II) rung-singlet regime for $0<J_\perp/J_\parallel \lesssim 5$, and (III) leg-singlet 
regime for $J_\perp/J_\parallel \gtrsim 5$.

This work was supported in part by the University of Tsukuba Research  
Initiative and KAKENHI 20029004, 20046002, 20654034 and 21740281 for M.A..

\end{document}